\documentclass[twocolumn,showpacs,preprintnumbers,amsmath,amssymb]{revtex4}

\usepackage{graphicx}
\usepackage{dcolumn}
\usepackage{bm}

\begin{document}

\preprint{APS/123-QED}

\topmargin 0pt

\title{A neurodynamic framework for local community extraction in networks} 

\author{Shihua Zhang$^1$}\email{zsh@amss.ac.cn}
\author{Guanghua Hu$^2$}
\author{Wenwen Min$^2$}%


\affiliation{%
$^1$National Center for Mathematics and Interdisciplinary Sciences,
Academy of Mathematics and Systems Science, Chinese Academy of
Sciences, Beijing 100190, China
$^2$Department of Mathematics, Yunnan University, Yunnan Kunming, 650091, China
}

%
%

\date{\today}

\begin{abstract}
To understand the structure and organization of a large-scale
social, biological or technological network, it can be helpful to
describe and extract local communities or modules of the network. In
this article, we develop a neurodynamic framework to describe the
local communities which correspond to the stable states of a
neuro-system built based on the network. The quantitative criteria
to describe the neurodynamic system can cover a large range of
objective functions. The resolution limit of these functions enable
us to propose a generic criterion to explore multi-resolution local
communities. We explain the advantages of this framework and
illustrate them by testing on a number of model and real-world
networks.

\end{abstract}

\pacs{Valid PACS appear here}


\maketitle

\section{Introduction}
In recent years, networks have emerged as an invaluable
tool to understanding systems of interacting objects in diverse
fields including sociology, biology and technology
\cite{Strogatz2001,Newman2003}. Due to the large-scale, complex
nature of many systems under study, one crucial step in network
analysis is the description and detection of mesoscopic structure
known as modules or communities: groups of nodes that are more
tightly connected to each other than they are to the rest of network
that have more internal links than external (see
ref.\cite{Fortunato2010} for a recent review). The network
communities form a distinct intermediate level and provide insight
into the structure of the overall network.

Traditionally, the community detection problem is formulated as
finding a ``best'' partition and multiple methods and heuristics
have been proposed for this issue \cite{Fortunato2010}. The
complexity of networks makes it complicated to measure the goodness
of a community structure and a variety of measures have been
proposed which have been shown to produce reasonable community
structure for a series of examples
\cite{Newman2004,Reichardt2006,Li2008,Bickel2009}. However, the
partitioning methods which force every node into a community can
distort the real structure of the network, in which, some nodes may
only loosely connected to any community. Moreover, the popular
measure modularity \cite{Newman2004,Newman2006} has been shown to
fail to find the most natural community structure due to the
resolution limit issues \cite{Fortunato2007} which leads to several
variants and extensions \cite{Muff2005,Reichardt2006,Li2008}.

In a large network, a community only focus on the ``local'' links
within it and links connecting it to a limited number of nodes of
the rest of the network. The principle of determining such a local
community at a time is different, but beneficial complement (view)
to the partitioning methods. There has been no much work in the
literature focusing on the local community detection. Researchers
have proposed local community methods aiming to look for the
community around a given node which relies on the predefined
knowledge \cite{Flake2002,Clauset2005}. In a very recent study, Zhao
\emph{et al.} \cite{Zhao2011} proposed a local community extraction
method by maximizing two quantitative measures via tabu search
technique. The resolution limit of the proposed criteria and
inefficiency of the local optimization technique partly inspire us
to explore further this issue.

\begin{figure*}[t]
\begin{center}
\centerline{\includegraphics[width=0.90\textwidth]{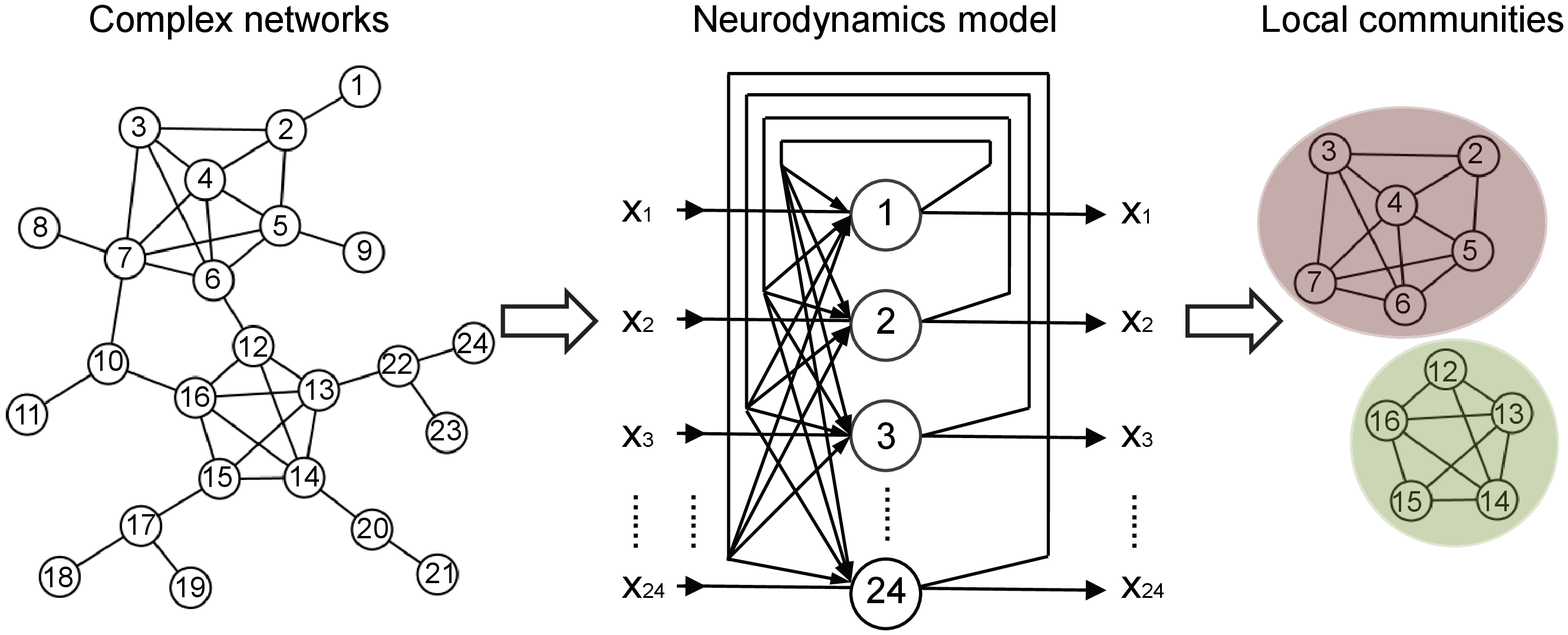}}
\caption{Illustration of the neurodynamic framework for modeling
local communities. (A) We want to describe the subnet of the network
such that tight substructures have distinct properties. (B) The key
to model the local communities is to build a proper neurodynamic
system $N(M,T)$ based on the structure of network $G(V,E)$, of
which, the stable states correspond to the local tight subunits. (C)
By running the system from a random initial state, the system can
converge to a stable state which corresponds to a local community as
we expected.}\label{Figure1}
\end{center}
\end{figure*}

In this article, we propose a neurodynamic framework to describe the
local organization of the links and nodes in networks that represent
the local dense subunits of a system (Figure \ref{Figure1}). The basic
idea is that local communities can be captured by making it
correspond to the stable states of a dynamic system. If one then
starts the dynamical system in random state that are sufficiently
close to one of the stable states, it should drift into one of these
stable states and stay there. We therefore identify the local
communities that compose the core part of a network by finding all
stable states of a neurodynamic system. Moreover, we can start the
system in a large number of random states, and then the frequency of
each stable state can shed light on the robustness of the
corresponding local community.


\section{Local community extraction problem}
Let $G(V,E)$ denote an undirected network of $n$ nodes; it can be
represented by a symmetric $n\times n$ adjacency matrix
$A=[A_{ij}]$, where $A_{ij} >0$ if there is an edge between nodes
$i$ and $j$ and $A_{ij} =0$ otherwise. If the edges have weights,
the positive $A_{ij}$'s are the weights; if not, they are set to 1.
The kernel idea of local community extraction problem is to look for
a set $S$ of nodes with a large number of links within itself and a
small number of links to the rest of the network. This problem can
be described to optimize a quantitative function, and note that the
links within the complement $S^c$ of this set do not affect the
value of this function. Here we employ two quantitative functions
$W_S$ and $Q_S$ to illustrate our framework and more other criteria
including the function for describing weak definition
\cite{Radicchi2004}, the minimum cut \cite{Shi2000}, the modularity
degree \cite{Li2008}, community density \cite{Bader2003} on
community can also be analyzed in the same manner (see Supplementary
file).

Here we introduced the $Q_S$ which is defined in sprit of the
modularity $Q$ \cite{Newman2004,Newman2006} and the $W_S$ adopted by
Zhao \emph{et al.} \cite{Zhao2011} as criteria for a local
community. Specifically,
\begin{equation}
Q_S=\frac{O_S}{O_V}-\left(\frac{O_S+B_S}{O_V}\right)^2, \label{qg1}
\end{equation}
and
\begin{equation}
W_S=|S||S^c| \left[\frac{O_S}{|S|^2}-\frac{B_S}{|S||S^c|}\right],
\label{qg2}
\end{equation}
where $O_S= \sum\limits_{i,j\in S} A_{ij}$, $B_S=\sum\limits_{i\in
S,j\in S^c} A_{ij}$. The term $O_S$ is twice the weight of the edges
(links) within $S$, and $B_S$ represents connections between $S$ and
the rest of the network. The first term of $Q_S$ is the fraction of
links inside community $S$ and the second term, in contrast,
represents the expected fraction of links in the community if links
were made at random but respecting node degrees in the network. The
first term of $W_S$ (in regards of the scalar $|S||S^c|$) is close
to the density of the community $S$, and the second term is expected
connections between the community and the rest of the network.
Intuitively, a local community should have high $Q_S$ and/or $W_S$.
Thus the task for resolving local communities can be changed into
find such `good' subnets by searching through the possible
candidates for ones with relatively high $Q_S$ and/or $W_S$.

In the following, we will show that the maximization of these two
criteria for resolving local community can be formulated into
integer quadratic (fractional) programming problems. Let's set
$$x_i=\left\{\begin{array}{cc}
               1, & i\in S \\
               0, & i\in S^c
             \end{array}
    \right.,
$$
then $|S|=\sum\limits_{i=1}^{n}x_i$,
$|S^c|=n-\sum\limits_{i=1}^{n}x_i$,
$O_S=\sum\limits_{i=1}^{n}\sum\limits_{j=1}^{n}A_{ij}x_i x_j$ and
$B_S=\sum\limits_{i=1}^{n}\sum\limits_{j=1}^{n}A_{ij}x_i (1-x_j)$.
Thus the maximization of the objective function $Q_S$ and $W_S$ can
be reformulated as the following optimization problems which both
subject to $x\in \{0,1\}^n$:
\begin{equation*}
\mbox{min}\quad -f_Q(x) =
-\frac{1}{2m}\sum\limits_i^n\sum\limits_j^n \left(A_{ij}-\frac{d_i
d_j}{2m} \right)x_i x_j,
\end{equation*}
\begin{equation}
=-\left(\frac{1}{m}\right)\frac{1}{2}x^TM_Q x. \label{qg3}
\end{equation}
where $M_Q=\left(A-\frac{d_id_j}{2m}\right)$, and
$$\mbox{min}\quad -f_W(x)=-\frac{\sum\limits_{i=1}^{n}\sum\limits_{j=1}^{n}A_{ij}x_i x_j (n-\sum\limits_{k=1}^{n}x_k) - \sum\limits_{i=1}^{n}\sum\limits_{j=1}^{n}A_{ij}x_i (1-x_j)\sum\limits_{i=k}^{n}x_k}{\sum\limits_{i=1}^{n}x_i}$$
\begin{equation}
=-(n)\frac{x^TM_Wx}{e^Tx}\equiv \frac{b(x)}{a(x)} \equiv \lambda(x),
\label{qg4}
\end{equation}
where $M_W=A-\frac{de^T}{n}$, $d=(d_1,d_2,\cdots,d_n)^T$,
$d_i=\sum_j A_{ij}$, $e=(1,\cdots,1)^T$. Since $x^TM_Wx=x^TM_W^Tx$
for any $M_W$, we can replace the $M_W$ with
$\frac{1}{2}(M_W+M_W^T)=A-\frac{de^T+ed^T}{2n}$. Then the new $M_W$
is a symmetric matrix 
which can ensure the convergence of the neurodynamic system
introduced later.

Generally, it is a challenging task to solve the quadratic
fractional programming (QFP) problem (Eq.4) directly. Fortunately,
according to Dinkelbach's theory \cite{Dinkelbach1967}, with $e^Tx >
0$, 
the fractional programming problem (Eq.\ref{qg4}) can be transformed into a
parametric programming model. This model will generate a sequence of
integer quadratic programs and the solutions of these programs can
converge to the solution of the fractional program. Specifically,
the problem Eq.(\ref{qg4}) can be transformed into the following
parametric quadratic programming problem:
\begin{equation}
\mbox{min}\quad g(x)=(2n)\left[-\frac{1}{2}x^TM_Wx-
\frac{\lambda}{2} e^Tx \right] \equiv b(x)-\lambda a(x), \label{qg5}
\end{equation}
where $b(x)=-x^TM_Wx$ and $a(x)=e^Tx$. By solving the quadratic
program with a given $\lambda$ and updating the
$\lambda^\prime=\frac{b(x^*)}{a(x^*)}$ with optimal solution $x^*$
of Eq.\ref{qg5} alternatively, we can finally obtain the solution of
Eq.(\ref{qg4}) (see Supplementary file for Dinkelbach theory).

We can see that the kernel problem for optimizing $Q_S$ and $W_S$ is
to solve the unconstrained Binary Quadratic Programming (BQP)
problem \cite{Kochenberger2004}. The BQP problem is an NP-hard
problem and has a large number of important applications in a broad
range of scientific fields. Various solution techniques including
both exact and heuristics have been proposed such as branch and
bound method, tabu search, simulated annealing and genetic algorithm
\cite{Merz2004}. However, due to the computational complexity of the
problem, most of these technique is limited for large-scale
problems.

\vspace{5ex}
\section{A neurodynamic framework}
We further mimic this problem with a neurodynamic system ---
Hopfield net model. As we analyzed that an `optimal' solution for
the above problem is a local stable dense region. Ideally, such a
solution can be related to a stable state of a Hopfield net system
which is built based on the topology of original network. In the
following, we will build the Hopfield net model for the local
community extraction problem. As we analyzed in the Supplementary
file that the Hopfield net model framework can cover a broad range
of quantitative functions, and we will further propose a generic
criteria (see \emph{A generic quantitative function} section) to
explore the complex hierarchical and multiple-resolution
characteristics which can also conquer the resolution-limit of
Eq.\ref{qg1} and \ref{qg2} (see \emph{Resolution limit analysis}
section).

Here we briefly describe the classic discrete Hopfield net model
\cite{Hopfield1982,Rojas1995}. Let's denote a discrete Hopfield net
of $n$ interconnected neurons as $H = (M,T)$, where $M$ is the
symmetric weight matrix of size $n\times n$ and $T$ is the threshold
vector of size $1\times n$. The neuron state vector is denoted as $x
= (x_1,x_2,\cdots,x_n)^T\in [0,1]^n$ and the neurodynamic system can
be described by:
\begin{equation}
x(t+1)=\textbf{sig}\{Mx(t)-T\},\label{qg6}
\end{equation}
where $\textbf{sgn}(x)=(\mbox{sgn}(x_1),\cdots,\mbox{sgn}(x_n))^T$
and the \mbox{sgn} is the signum function defined as $\mbox{sgn}(x)$
equal to 1, if $x\geq0$, otherwise 0. The well-known fundamental
property of this system is that their dynamics are constrained by an
energy function $E(\textbf{x})$ (also known as Lyapunov function)
defined on their state space by:
\begin{equation}
E(\textbf{x})=-\frac{1}{2}\textbf{x}^TM\textbf{x}+T^T\textbf{x}.
\label{qg7}
\end{equation}
If the system state converges toward some stable state, then this
stable state correspond to a local minimum of $E(\textbf{x})$. It
has been proved that the asynchronous Hopfield system starting from
any initial state converge to a stable state provided that
$w(i,i)\geq 0$ for all $i$, while the synchronous one converge to a
stable state or to a limit cycle of length two under
mild hypotheses 
\cite{Sima2003}.

We can easily see that the formulaic form of BQP problem is
isomorphic to the energy function of Hopfield net, therefore the
local community extraction problem defined based on $Q_S$ and $W_S$
can be solved by the neurodynamic system. The components of Hopfield
nets of them can be defined as $M_Q=A-\frac{d_id_j}{2m}$, $T_Q=0$
and $M_W=A-\frac{de^T+ed^T}{2n}$, $T_W=-\frac{\lambda}{2}$
respectively (we can omit the scalars). The topology of this system
has very natural corresponding relationship with the original
network. Particularly, the sparsity of the original network can be
employed to accelerate the dynamical update. Moreover, the
neurodynamic framework enables us to study the robustness of the
local community structure based on the properties of its
corresponding stable state.

\section{Algorithm}
The synchronous discrete Hopfiled system can be run in a more
efficient manner than the asynchronous one, so we adopt it as the
basic procedure to solve the BQP problem. The Synchronous Discrete
Hopfiled Net (named SDHN) can be implemented as follows: \\
\textbf{SDHN$\{H(M,T)\}$}
\begin{itemize}
  \item Step \textbf{1}: Select an initial vector $\textbf{x}(0)\in [0,1]^n$, and set $t = 0$;
  \item Step \textbf{2}: Update $\textbf{x}(t)$ by $\textbf{x}(t+1)=\textbf{sig}\{M\textbf{x}(t)-T\}$;
  \item Step \textbf{3}: If $\textbf{x}(t + 1)$ satisfies a given stop criterion, then stop and output $\textbf{x}(t + 1)$; else $t := t + 1$ and go to Step
  \textbf{2}.
\end{itemize}

We further employ the SDHN algorithm as subroutine to solve the QFP as follows:  \\
\textbf{QFP$(a(x),b(x))$}
\begin{itemize}
  \item Step \textbf{1}: Select an initial vector $\textbf{x}(0)\in [0,1]^n$, and set $\lambda(0)=\frac{a(\textbf{x}(0))}{b(\textbf{x}(0))}$ and
  $k=0$.
  \item Step \textbf{2}: Solve Eq.(\ref{qg5}) using SDHN$\{H_k(M,T_\lambda)\}$ to get the stable state
  $\textbf{}x(k+1)$.
  \item Step \textbf{3}: If $a(\textbf{x}(k+1))-\lambda(x_k)b(\textbf{x}(k+1))=0$), then set
  $x^*=x(k+1)$ and $\lambda^*=\lambda(k)$, and stop.
  \item Setp \textbf{4}: If $a(\textbf{x}(k+1))-\lambda(x_k)b(\textbf{x}(k+1))>0$),
  then set $\lambda(k+1)=\frac{a(\textbf{x}(k+1))}{b(\textbf{x}(k+1))}$,
  $k=k+1$ and go to Step \textbf{2}.
\end{itemize}

Since the topology $W$ of the Hopfield system only related to the
adjacency matrix $A$ of a network or $A+B$ where $B$ is usually of
rank 2, then neurodynamic system can be updated in $O(m+n)$, where
$m$ is the numbers of edges and $n$ is the number of nodes in the
network. The $\lambda$ can also be efficiently updated in $O(m+n)$.
If we take into account the number of iterations $L$, we have the
computation effort $O(L(m+n))$ for each run of the neurodynamic
system (SDHN procedure). We observed that this procedure usually
converges in a small number of iterations (e.g., $L=10$). More
importantly, we only need to update the $\lambda$ for \textbf{about
several times} to converge in running the QFP procedure. Thus the
whole neurodynamic framework can be run in a near linear time cost
for one trial. Although the neurodynamic system may be trapped in
the trivial stable state ($x=[0,0,\cdots,0]^T$), the high efficiency
of this framework makes it possible to run a proper number of trials
(e.g., 500). Our neurodynamic procedure is an very efficient
algorithm which enable it to be a powerful way to extract the local
communities in a large-scale network. While the stochastic search
methods like tabu search approach are highly time-consuming
techniques which can only be applied to network of at most a few
thousand vertices with common hardware.

After determining a local community, we can further apply the whole
framework to its complement in the network to extract next
community. A challenging and open problem faced by all methods for
community-detection is how to determine the number of communities in
a network. In real applications, we would suggest to evaluate the
statistical significance of a community by comparing its objective
value with that of 100 random networks generated by containing the
same set of nodes and the same number of edges
\cite{Maslov2002}.

\section{Resolution limit analysis}
Unfortunately, due to the improper penalty concerning to the total
size of networks, this local modularity criterion has serious
resolution limit as found for the popular modularity function
\cite{Newman2004,Newman2006}. This resolution limit problem of
modularity has been carefully discussed by Fortunato and Barthelemy
\cite{Fortunato2007}. To illustrate the resolution limit of the
local modularity function, we analyze the local modularity of
`communities' in several schematic examples. The ring-clique network
consisting of a ring of cliques and connecting through single links
adopted by Fortunato and Barthelemy in ref. \cite{Fortunato2007} is
employed (Figure 2A); each clique is a complete graph $Km$ with $m$
nodes and has $m(m- 1)/2$ links. If we assume that there are $n$
cliques (with $n$ even), the network has a total of $N= nm$ nodes
and $L= nm(m-1)/2+ n$ links. Obviously, this network has a clear
community structure where the communities correspond to single
cliques, and we expect that any community extraction algorithm
should be able to detect these communities.

The local modularity $W_{clique}$ of a clique can be easily
calculated and is equal to
$$W_{clique} =m(m- 1)(n-1)-2.$$
On the other hand, the local modularity $W_{pairs}$ of the pairwise
consecutive cliques are considered as single communities (as shown
by the dotted lines in Figure 2A) is
$$W_{pairs} = m(m- 1)(n-2)+n-4.$$
The difference $\bigtriangleup W=W_{clique} -W_{pairs} $ is
$$\bigtriangleup W=m(m-1)+2-n,$$
Then the condition $W_{clique} > W_{pairs}$ is satisfied only if
\begin{equation}
m(m-1)+2>n,  \label{qg8}
\end{equation}
Obviously, $m$ and $n$ are independent variables, and we can choose
them such that the inequality of Eq.\ref{qg8} is not satisfied. For
instance, for $m= 10$ and $n = 100$, $W_{clique}=8908$ and
$W_{pairs}= 8916$. Then an efficient algorithm would find the
configuration with pairs of cliques and not the single clique as an
`ideal' community to get the maximum local modularity. As we can see
that, as $n$ increases, the difference $-\Delta W$ becomes even
larger.

This situation can also be observed in another example (shown in
Figure 2B), where the two pink circles again represent cliques with $p$
nodes linked each other by a single edge, the other part is the rest
of the network with $n$ nodes linked these two cliques by single
link respectively. We can similarly calculate the local modularity
of a clique as $W_{clique}=(n+p)(p-1)-2$ and the pair cliques as
$W_{pairs}=n(p-1)+n/p-2$. Then the difference $\Delta W=W_{clique}
-W_{pairs}$ is $\Delta W=\frac{p^2(p-1)-n}{p}$. So if $p^2(p-1)<n$,
the two cliques will merge as one community. For example, if we take
the network of Figure 2B with $p=10$ and $n=1000$ and the $n$ nodes was
randomly connected with probability $p=0.05$, we have seen that the
local modularity criterion will get a``module" which are pairs of
connected cliques. Although the example was very simple, we can
clearly see that the local community was only affected by the number
of nodes in the rest of the network, while not the specific
structure of it. This means that, in large-scale sparse networks,
the local dense communities are easily to be merged together which
was also further observed in our simulation study as well as in real
networks.
\begin{figure}[t]
\begin{center}
\centerline{\includegraphics[width=0.47\textwidth]{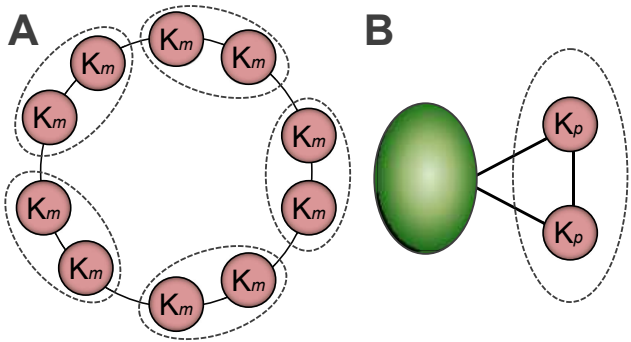}}
\caption{Illustrative examples. (A) A network composed of identical
cliques with $m$ nodes connected by single links. If the number of
cliques $n$ is larger than about $m(m-1)+2$, the local modularity
optimization would lead to a local community which are  combined of
two or more cliques (represented by dotted lines). (B) A network
with two identical cliques of $p$ nodes linked by a single link
between them, and other subnetwork of $n$ nodes with arbitrary
connections; the two cliques are linked to the subnetwork with a
single link respectively. If $n$ is large enough with respect to $p$
(e.g., $n=1000$, $p=10$), the local modularity optimization merges
the two cliques into one (shown with a dotted line).}
\label{Figure2}
\end{center}
\end{figure}

\section{A generic quantitative function}
As discussed by Zhao et al. \cite{Zhao2011}, the factor
$|S||S^\rho|$ in $W_S$ tend to extract larger communities and avoid
the smaller communities without this factor. However, this factor
relate a local community with the size of network (similar to $Q$
related with number of total links) which further lead to the
resolution limit. Taking into account the resolution limit of $W_S$,
we propose a parametric quantitative criterion to extract local
communities and explore multi-resolution community characteristic of
networks. The generic quantitative function $W_S^{\rho}$ is defined
as follows by introducing a parameter $\rho$:
\begin{equation}
W_S^{\rho}=|S||S^\rho|
\left[\frac{O_S}{|S|^2}-\frac{B_S}{|S||S^\rho|}\right], \label{qg9}
\end{equation}
where $|S^\rho|=\rho n-|S|$, $2\frac{|S|}{n}<\rho\leq 1$. The
$|S^\rho|$ can be considered as the estimation of the number of
nodes connecting to the community $S$ in the rest of the network.
\textbf{[}$2\frac{<S>}{n}<\rho\leq 1$, and $<S>$ is the expected
number of community size. Then if the factor $|S|=\frac{\rho n}{2}$,
$|S||S^\rho|$ gets its maximum, i.e.,
$\rho=\frac{2|S|}{n}$.\textbf{]} If $\rho=\frac{2|S|}{n}$, the
criterion is become $O_S-B_S$ which is related to the weak
definition \cite{Radicchi2004}, while if $\rho=1$, the criterion is
just the $W_S$. Similar with $W_S$, the generic function
$W_{S}^{\rho}$ can be solved by the topology-varying neurodynamic
model efficiently. In particular, the components of corresponding
Hopfield net can be defined as $M_{W^\rho}=\rho
A-\frac{de^T+ed^T}{2n}$, $T_W=-\frac{\lambda}{2}$ (we can omit the
scalars). In real applications, we can sample $\rho$ from the given
range to study the detailed community structure of networks
systematically.


\section{Simulation}
We generate a network consisting two tight communities and weakly
connected background as suggested by Zhao \emph{et al.}
\cite{Zhao2011}. We consider community sizes $n_1=100?$ and
$n_2=200?$, which are embedded into the background nodes forming the
network of size $n$ ($n=1000$ and $10000$). The links between a
community, and the links between members and others or links between
nodes in the background all form independently with probability 0.3
and 0.05. We compare the modularity partition (with spectral
optimization \cite{Newman2006}) and the local community extraction
with $W_S (\rho = 1)$ and $W_S^\rho=0.6$. We partition the network
into three parts by modularity and extract two communities by the
extraction methods. We show the accuracy to compare these networks.

We should note that the parameter is tunable and we can obtain a
spectrum of `local` communities by sampling $\rho$ in the range
$[\rho_{min},1]$. Such spectrum can shed light on the underlying multiple
resolution or hierarchical community structure of networks. To show
this point, we simulated a network of size $n=1000$ with multiple
resolution community structure. The spectrum constructed
by our method with different $\rho$ can well uncover such
complicated communities in the network. We believe this
will benefit the understanding of the underlying mesostructure of
networks.

We have shown that the neurodynamic framework can accurately
identify the communities in the simulated networks with
significantly higher precision than that of the original one and the
modularity, and the generic quantitative criterion can conquer the
resolution limit of the original one. As we have analyzed that the
cost of the neurodynamic procedure is only about $O(T(m+n))$. Here
the $T$ is about 10-20 for network with $n$ between 100 to
10000. We further estimate how many trials must be used to
identify the optimal community. We found 500 random trials are
enough to get the best solution. We have compare the time cost of
our procedure with the tabu search for maximize $W_S$. The
tabu search can only be applied to the network with thousands of nodes,
while our procedure can work on the network with 20000 nodes in
30 seconds. The efficiency of our method make it feasible to real
large-scale networks.

\begin{figure}[t]
\begin{center}
\centerline{\includegraphics[width=0.48\textwidth]{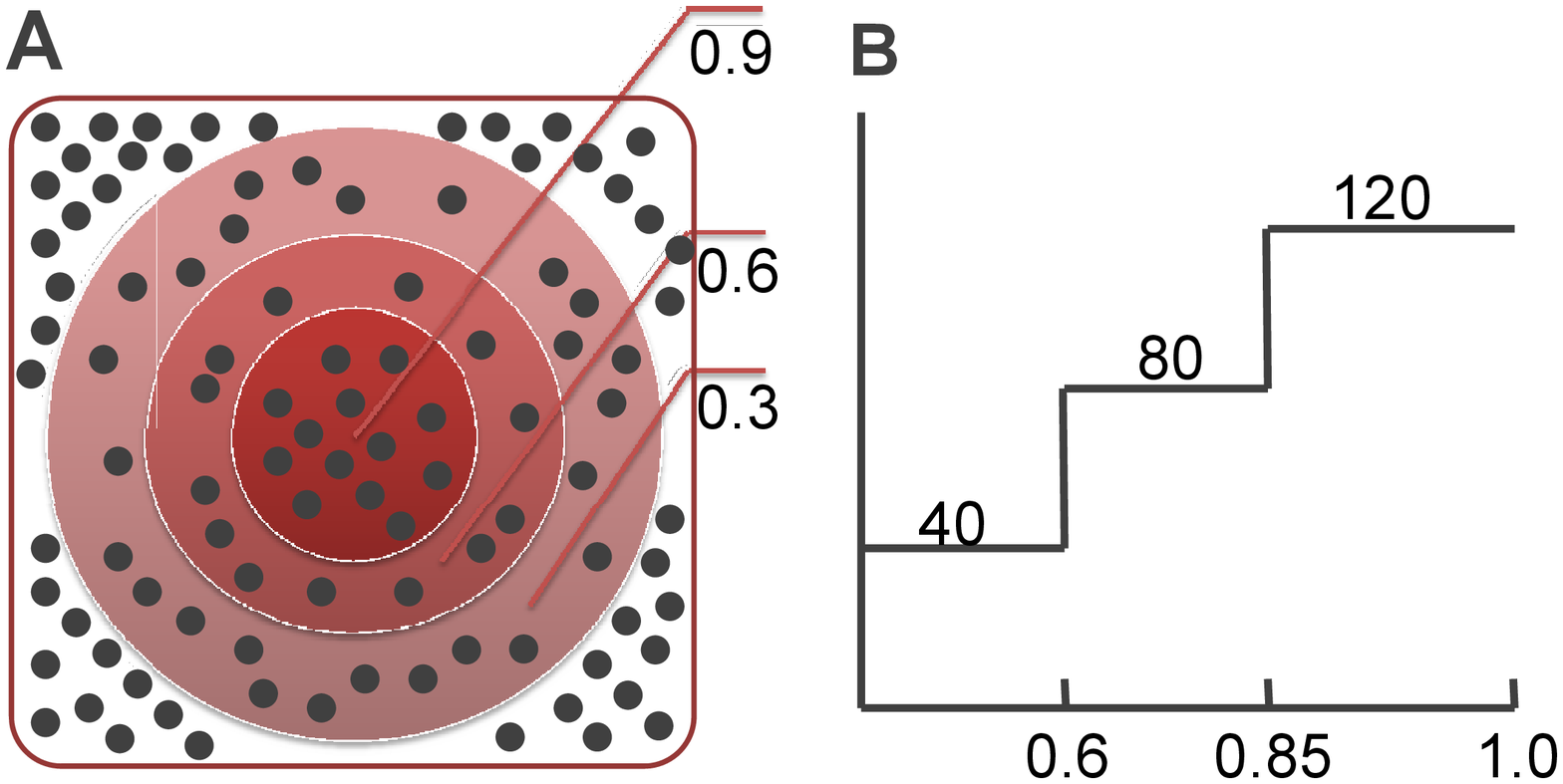}}
\caption{(A) The local communities identified by our approach in the
simulated network with multiple-resolution community structure.
Different colors illustrate the membership of each local community.
(B) The membership of local communities identified by our approach
with different $\rho$.} \label{Figure3}
\end{center}
\end{figure}

\section{Example Applications}
\begin{figure}[t]
\begin{center}
\centerline{\includegraphics[width=0.40\textwidth]{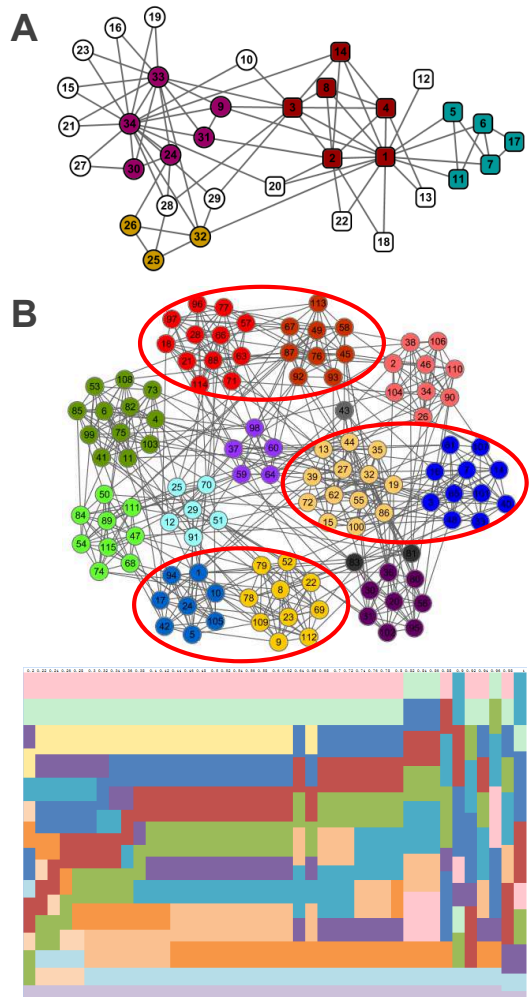}}
\caption{(A) The local communities identified by our approach in the
karate club network. Shapes of nodes indicate the membership of the
corresponding individuals in the two known factions of the network.
Different colors illustrate the membership of each local community.
(B) The local communities identified by our approach in the football
network with $\rho=0.4$ and $\rho=1$ and the membership of local
communities identified by our approach with different $\rho$.}
\label{Figure4}
\end{center}
\end{figure}

\textbf{Two well-known examples with significant community
structure.} The first example is the well-known network of
friendship between 34 members of a karate club at an American
university \cite{Zachary1977}. This network is of particular
interest because the club split into two subclubs due to an internal
dispute, and the structure of the recorded network reflects the
trend of this division. The second example is the network
representing the schedule of Division I football games for the 2000
season \cite{Newman2004}. The network consisting of 115 nodes and
613 links representing teams and regular season games between two
teams, are divided into conferences with more games are played
within conferences than across them.

The partitioning methods including modularity can partition these
two networks into (almost) the exact known groups \cite{Newman2006}.
However, the local community extraction method reveals the `core'
structural information about these (this) networks. Figure \ref{Figure4}A shows
the karate club network and its local communities extracted by our
method with $\rho =1$, which are consistent with the results
obtained by optimizing the criterion using taboo search technique
\cite{Zhao2011}. We can observe that the first two communities
occupy the core part of the original two groups which determine the
evolving trend, and the third one is a subcommunity in the
\textbf{instructor group} which is prone to a further division. We
should note that although the order of objective values have
changed, we can get the same local communities with $\rho$ ranging
from 0.6 to 1 (Figure \ref{Figure4}B).


However, when we apply the $W_S$ (i.e., $W_S^\rho$ with $\rho=1$) to
the football network, we can observe clear resolution limit problem.
Figure \ref{Figure4} show the extracted communities with $\rho=1$ and
$\rho=0.4$. The local communities identified with $W_S$ tend to
combine several groups together which clearly show the resolution
limit of the quantitative function $W_S$. While our neurodynamic
system based on the generic function with $\rho=0.4$ can well
identify those local communities that correspond to the \textbf{core
part of} real groups well.

\section{Conclusion}
We have shown that the process of resolving local communities in
complex networks can be viewed as finding stable states in a
neurodynamic system. By describing local communities with a
quantitative function, we are able to construct a corresponding
neurosystem which can store the structural patterns as stable states
among it. The solution of optimizing the quantitative function for
describing local communities is, in principle, NP-hard. However, the
neurodynamic system can be run in an very efficient manner which
makes it can applicable to large-scale networks, while the heuristic
search method can not (e.g., tabu search used by ref.
\cite{Zhao2011}). The original local community criterion proposed by
Zhao \emph{et al.} \cite{Zhao2011} was also hindered by serious
resolution limit issue as showed for modularity
\cite{Fortunato2007}. Actually, as we stated that any quantitative
function related to the whole size of network (the total number of
nodes or links) will bear this problem. A generic parametric
quantitative function is in demand which can help to explore the
complicated multi-resolution structure of networks. We believe that
the local community extraction framework proposed here is an
important complementary to the traditional partitioning methods
which may bear distinct underlying problem due to the global modular
hypothesis. We expect the proposed method will benefit network
science with broad applications in various fields including biology,
sociology and technology.

\begin{acknowledgments}
This work was supported by the National Natural Science Foundation of China, No. 61379092, 61422309
and 11131009, the Strategic Priority Research Program of the Chinese Academy of Sciences (CAS) (XDB13040600),
the Outstanding Young Scientist Program of CAS and the Key Laboratory of Random Complex
Structures and Data Science, CAS (No. 2008DP173182).
\end{acknowledgments}

\end{document}